# Fully three-dimensional sound speed-corrected multi-wavelength photoacoustic breast tomography

**M. Dantuma[1], F. Lucka[2,3], S. C. Kruitwagen[4], A. Javaherian[5], L. Alink[6], R. P. Pompe van Meerdervoort[6], M. Nanninga[6], T.J.P.M. Op 't Root[6], B. De Santi[1], J. Budisky[7], G. Bordovsky[7], E. Coffy[8], M. Wilm[8], T. Kasponas[9], S.H. Aarnink[4], L. F. de Geus-Oei[10,11], F. Brochin[8], T. Martinez[8], A. Michailovas[9,12], W. Muller Kobold[6], J. Jaros[7], J. Veltman[1,13], B. Cox[5, ✉], and S. Manohar[1, ✉]**

[1] Multi-Modality Medical Imaging group, TechMed Centre, University of Twente, Enschede, The Netherlands

[2] Computational Imaging, Centrum Wiskunde & Informatica, Amsterdam, The Netherlands

[3] Department of Computer Science, University College London, London, United Kingdom

[4] Medisch Spectrum Twente Hospital, Enschede, The Netherlands

[5] Department of Medical Physics and Biomedical Engineering, University College London, London, United Kingdom

[6] P.A. Imaging R&D B.V., Enschede, The Netherlands

[7] Department of Computer Systems, Faculty of Information Technology, Brno University of Technology, Brno, Czech Republic

[8] Imasonic SAS, Voray-sur-l'Orgnon, France

[9] Ekspla UAB, Vilnius, Lithuania

[10] Biomedical Photonic Imaging group, University of Twente, Enschede, The Netherlands

[11] Department of Radiology, Leiden University Medical Center, Leiden, The Netherlands

[12] Center for Physical Science and Technology, Vilnius, Lithuania

[13] Department of Radiology, Ziekenhuisgroep Twente, Hengelo, The Netherlands

✉ s.manohar@utwente.nl; ✉ b.cox@ucl.ac.uk

**Photoacoustic tomography is a contrast agent-free imaging technique capable of visualizing blood vessels and tumor-associated vascularization in breast tissue. While sophisticated breast imaging systems have been recently developed, there is yet much to be gained in imaging depth, image quality and tissue characterization capability before**

**clinical translation is possible. In response, we have developed a hybrid photoacoustic and ultrasound-transmission tomographic system (PAM3). The photoacoustic component has for the first time three-dimensional multi-wavelength imaging capability, and implements substantial technical advancements in critical hardware and software sub-systems. The ultrasound component enables for the first time, a three-dimensional sound speed map of the breast to be incorporated in photoacoustic reconstruction to correct for inhomogeneities, enabling accurate target recovery. The results demonstrate the deepest photoacoustic breast imaging to date (48 mm),with a more uniform field of view than hitherto, and an isotropic spatial resolution that rivals that of Magnetic Resonance Imaging. The *in vivo* performance achieved, and the diagnostic value of interrogating angiogenesis-driven optical contrast as well as tumor mass sound speed contrast, gives confidence in the system's clinical potential.**

Photoacoustic (PA) tomography, is a non-invasive imaging modality that relies for physical contrast predominantly on differential optical absorption of tissue constituents[1,2]. Unlike in purely optical techniques where light that survives the absorption is measured for the signal, in PA a surrogate of the actual absorbed nanosecond pulse optical energy is detected, namely acoustic waves[3]. These waves in the ultrasound (US) frequency regime are launched by transient thermoelastic expansion when the absorbed optical energy is converted into heat by non-radiative deexcitation processes. Since the propagating US wavefronts in soft tissue experience relatively low scattering compared to light, their faithful detection at the tissue surface enables accurate localization of the often deeply located US sources. The method is thus attractive since it combines the ability to interrogate intrinsic optical spectroscopic contrast of tissue components, with the ability to localize this contrast deep inside soft tissues with high resolution[4-8].

PA tomography is interesting for investigating the female breast due to the organ's relatively low acoustic scattering and absorption, as well as its accessibility for light excitation and US detection from all directions when pendant. Importantly the breast is the site of the highest incidence of female cancers globally, with 2.3 million new cases and 685,000 deaths in 2020[9], while established breast imaging methods suffer from drawbacks and limitations. X-ray mammography uses ionizing energy, applies often painful breast compression, and shows poor performance in visualizing cancer in breasts rich in fibroglandular tissue[10-12]. US imaging has poor discrimination between malignancies and benign abnormalities in the evaluation of solid masses[11,13]. Magnetic Resonance Imaging (MRI) shows excellent sensitivity by mapping contrast agent kinetics in tumor vasculature, but there is considerable overlap in this behavior between malignancies and benign abnormalities[11,14]. The method also requires contrast agents, is expensive, and not universally available.

PA tomography uses non-ionizing radiation, has low burden to the patient and has been demonstrated to be able to image blood vessels associated with breast cancer with high resolution, without the use of contrast agents[15]. Various PA breast imaging configurations have been developed, determined by the US detection aperture, and can be roughly divided into four geometries: linear, planar, curved/circular and hemispherical[15]. Of these, the last mentioned is most appealing because the US detectors can capture the PA emissions over a solid angle of $2\pi$ steradians, which gives sufficient data to reconstruct any object within the hemisphere without limited view artifacts[16,17].

The latest and most sophisticated hemispherical geometry PA breast imager was presented in 2021 by Lin et al[5]. The imager, using a single wavelength of 1064 nm and US detection with 1024 elements, revealed detailed vascular anatomy in the breast with an isotropic spatial resolution of 0.37–0.39 mm, to depths with a maximum of 40 mm and with an imaging time of 10 seconds per breast[5]. The system represents the state-of-the-art in breast PA computed tomography but still employs non-uniform illumination at a single wavelength, which impacts on the accuracy of imaging and the ability to extract quantitative spectral information. Further, this imaging system and its image reconstruction, as all others before it[18-21], treat the breast as being acoustically homogeneous.

In this paper, we present a hybrid breast imaging system (PAM3) that has for the first time multi-wavelength PA imaging capability, as well as the capability to measure a full three-dimensional (3D) sound speed map of the breast through hybrid ultrasound-transmission tomography (UST). The volumetric sound speed distribution in the acoustically complex breast is used to correct the wavefront aberrations experienced by the PA waves as they traverse the breast. This enables highly accurate recovery of the PA sources that leads to deeper imaging than hitherto possible. A standard image acquisition protocol uses 5 wavelengths and takes 5 minutes. The PA tomography mode is a major technological improvement in itself due to advancements in sub-systems such as the laser, light delivery system, ultrasound transducers, electronics and image reconstruction algorithms. We present details of the PAM3 system, a comprehensive characterization of its capabilities on a specially-developed suite of test-objects and sophisticated phantoms, and demonstrate its performance on healthy breasts.

# Results

## System description

The PAM3 system is built into a custom-designed frame with bed-top (2.5 m x1 m size) for the woman to lie on prone, with one breast through the imaging aperture which leads into a water-filled imaging bowl beneath. Figure 1a is an illustration of the system with a side panel removed to show the hardware groups under the bed-top containing respectively: i) the laser system, ii) the rotating core of the system, and iii) the water-temperature-control system. The top-view of the imaging bowl, as seen through the aperture, is shown in the inset of Fig. 1a. Supplementary Fig. A1 shows photographs of the system and the imaging bowl. The hemispherical bowl has a 26 cm inner diameter and has holes arranged in a spiral fashion to accommodate the following water-tight inserts: 40 optical fiber bundle terminations to provide light for PA excitation, 512 single-element transducers for the detection and emission of ultrasound, and two platinum resistance thermometers to measure water temperature. Each fiber bundle termination (Fig. 1b) comprises a 20 mm length tapered silica waveguide to homogenize the beam, followed by a fused silica lens array to shape the beam into a top-hat profile and ensure that the beam diverges rapidly to achieve a good coverage of the breast. The US transducers (Fig. 1b) (Imasonic, Voray-sur-l'Ognon, France) have a 1 MHz center frequency, and each comprises a 3 mm diameter piezocomposite active element which can be used both in detection and emission modes.

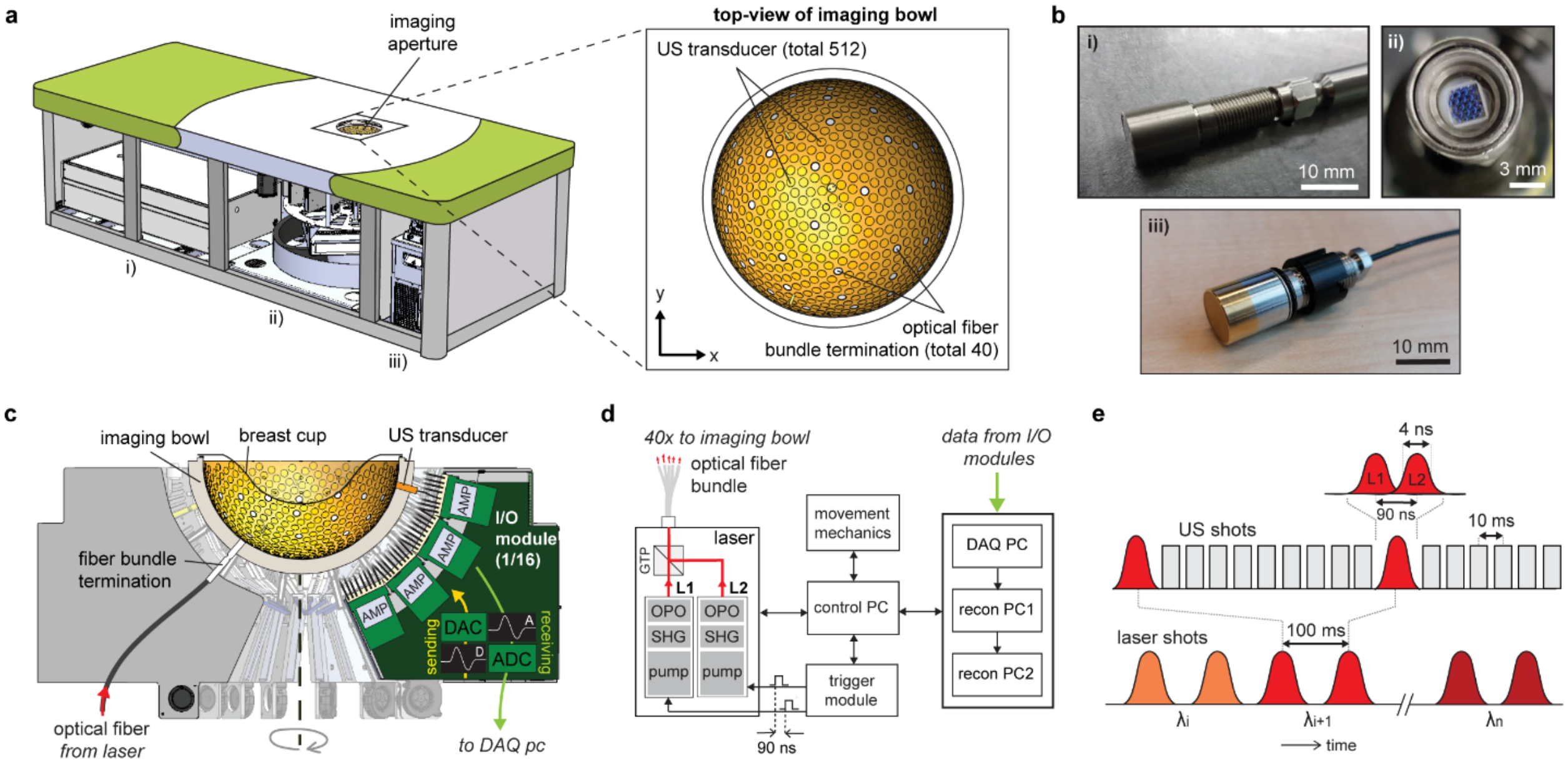

**Figure 1.** Overview of the PAM3 breast imaging system. The subject lies in a prone position with one breast through the imaging aperture into the temperature-stabilized water-filled imaging bowl. The bowl has two sets of bowl-inserts namely ultrasound (US) transducers and optical fiber bundle terminations. **a,** Illustration of the system in perspective view with one of the side panels removed to reveal three hardware compartments containing i) the laser, ii) the rotating turret and iii) temperature control unit. Inset is the top-view of the imaging bowl showing the arrangement of the bowl-inserts. **b,** Photographs of i) a fiber bundle termination, ii) the output microlens array termination of an optical fiber bundle, and iii) an US transducer. **c,** Cross-sectional view of the core of the system, showing the imaging bowl, and I/O modules arranged around it. There are in total 16 I/O modules each comprising 32-channel Analog-Digital converters (ADC) and Digital-Analog converters (DAC) to service the 512 US transducers. **d,** Block-scheme showing various modules, their interconnections and signal transfers. **e**, timing diagram of laser and US firing for a measurement sequence consisting of n wavelengths and two PA averages. AMP = pre-amplifier, recon = reconstruction, DAQ = data acquisition, L1 = laser head one, L2 = laser head 2, OPO = optical parametric oscillator, SHG = single harmonic generator, GTP = Glan-Taylor polarizer.

Figure 1c shows a cross-section of the rotating core of the PAM3 system, and Fig. 1d is a block-diagram showing various modules with their coupling and data transfers. An optically and acoustically transparent polyvinyl chloride (PVC) cup[22] is used to centralize and stabilize the breast in the bowl, and mitigate motion of the breast during a measurement. The breast-supporting PVC cups have an approximate thickness of 180 $\mu$m[22], with holes to ensure that the breast is in contact with water for good acoustic coupling with the US transducers.

The 40 optical fiber bundles are brought together and coupled to the PA excitation source (Ekspla UAB, Vilnius, Lithuania) which comprises two identical units each consisting of a Q-switched Nd:YAG pump laser, a DKDP crystal second harmonic generator (SHG), and a BBO crystal optical parametric oscillator (OPO). In normal operating mode, the pump lasers are triggered with a 90 ns delay between them, to avoid overlap of the two pulses in time in order to stay under the damage threshold of the fiber bundle (see Fig. 1e). The two outputs are combined using a Glan-Taylor prism (GTP) into single beam. The US transducer cables lead to input-output (I/O) electronic modules. During a measurement, the bowl together with the inserts, their connections and couplings, and the I/O modules, rotates step-wise over 360 degrees to acquire multiple-views around the breast. The spiral arrangement of the US transducers ensures that their positions remain unique during a full rotation. Each I/O electronics module (PA Imaging R&D B.V., Enschede, The Netherlands) services 32 US transducer elements and comprises a Data Acquisition (DAQ) part and an US Actuation part. The system control software allows individual measurement sequences to be programmed (number of bowl rotational steps, number of wavelengths, number of averages, number of US shots). The timing diagram of an example measurement sequence consisting of 5 wavelengths, two PA averages and 9 US shots is illustrated in Fig. 1e. This sequence is repeated at each bowl rotational position.

**Light delivery, ultrasound detection and breast supporting cups**

The output energy of the PA excitation system varies from 450 mJ at 680 nm to 230 mJ at 1060 nm (Fig. 2a). This combined energy is roughly 3 times higher, and each laser-OPO's energy 1.5 times higher, than the reported energy from lasers in the same class. The inset shows the output intensity profile at 800 nm of one of the 40 optical fiber bundles measured at 23 mm distance in air with a beam profiler (DataRay Inc, Redding, USA). In the imaging bowl, the beam radiates into the water with an opening angle of 55 degrees. The arrangement of the fiber bundle endings and the high beam divergence in water provides a homogeneous illumination profile on the breast surface while keeping the radiant exposure on the breast below the Maximum Permissible Exposure (MPE) for skin (Supplementary Fig. A2). When an appropriately-sized breast-supporting cup is selected from the eight sizes available (Fig. 2b), for a measurement on a human subject, the breast fills the cup and takes its shape. The light distribution on the breast surface can be therefore visualized with PA measurements on light absorbing breast-supporting cups made from 1 mm thick black unplastified PVC. The uniform appearance of the PA reconstructions (Fig. 2c) as maximum intensity projections (MIPs) of measurements on different cup sizes demonstrates the homogeneity of radiant exposure on the cups and thus the breast (See also Supplementary Table A1.). The US transducers were measured to have an average minimal detectable pressure (MDP) of 0.30 Pa, the lowest reported in the literature[23], with a standard deviation of 0.02 Pa (Fig. 2d). The center frequency lies around 1 MHz with a 123% fractional frequency bandwidth (Fig. 2e).

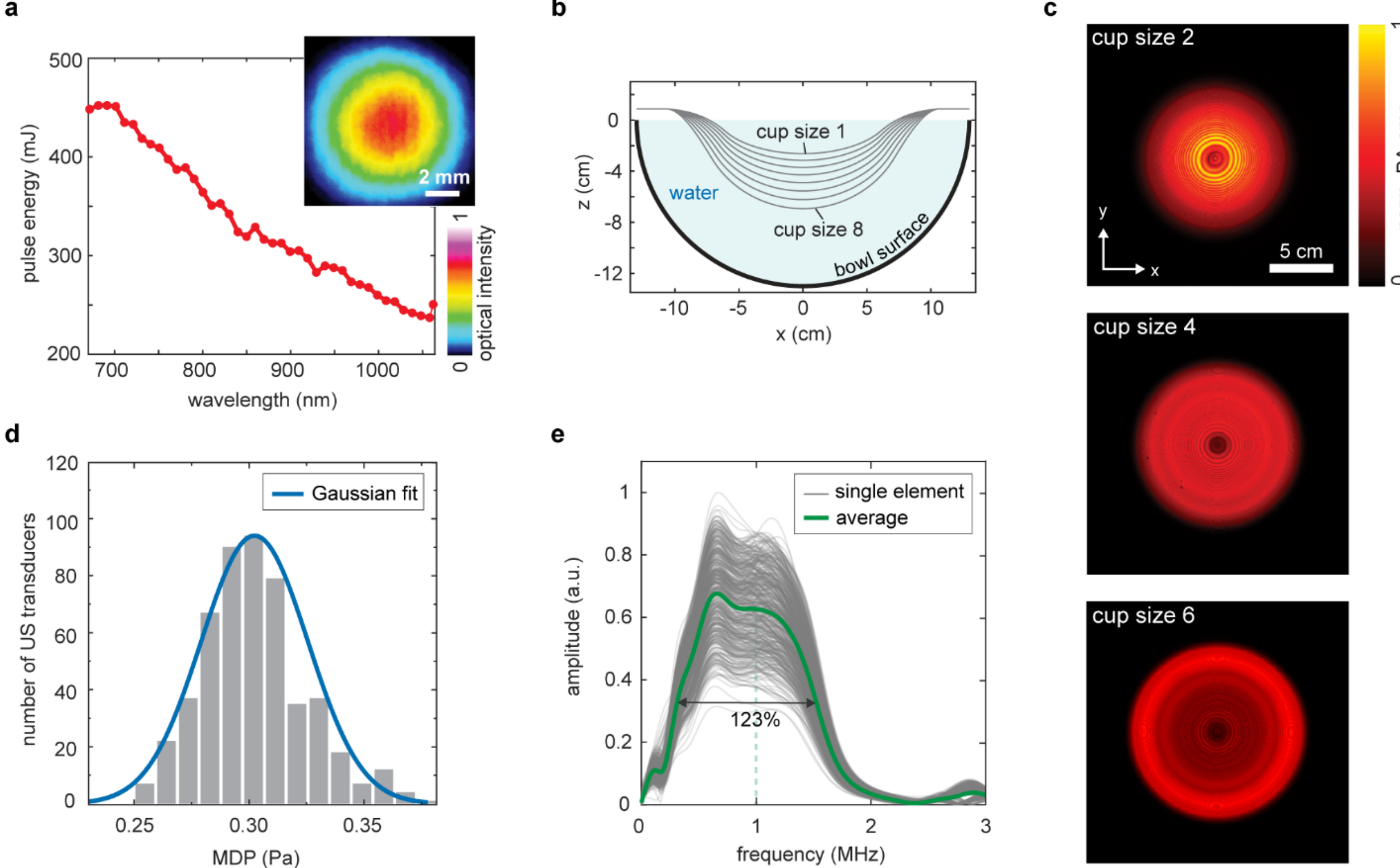

**Figure 2.** Characteristics of the custom-developed laser unit, light delivery approach, US transducer elements, and breast-supporting cups. **a,** The laser pulse energy as a function of wavelength. The inset shows the beam profile at 800 nm at a distance of 23 mm in air from one of the 40 fiber bundle terminations. **b,** Radially symmetric contours through the centres of the eight breast-supporting cups. **c,** Maximum intensity projections (MIP) of PA reconstructions of a selection of 3 black breast-supporting cups in the imaging bowl. The uniform reconstructed intensity shows that the cups and thus the breast surface receives a homogeneous radiant exposure. **d,** Histogram of the minimum detectable pressure (MDP) distribution of the 512 US transducers. A Gaussian fit of the data gives a mean MDP at 0.3 Pa. **e,** The measured frequency responses of the 512 US transducers showing an average fractional frequency bandwidth of 123% around 1 MHz.

**Test objects and phantoms imaging**

The imaging performance of the PAM3 system was assessed and optimized through a series of measurements on a specially developed suite of test objects and phantoms[24-26]. The 3D spatial resolution was assessed by imaging a test object consisting of multiple sub-resolution (± 300 $\mu$m diameter) PA targets[24] embedded in polyvinyl chloride plastisol (PVCP) in a size 8 breast cup (Supplementary Fig. A3). Figure 3a shows the MIP along the *x*-axis of the PA reconstruction of the test object, for 101 bowl rotational steps and 10 reconstruction iterations. The PA image reconstruction method using an accelerated, iterative, first-order optimization procedure[27] can utilize the measured sound speed map in PA reconstruction, as well as take into account the spatial and temporal detector responses. The dashed box marks a selected point target of which a zoomed-in view is shown. Increasing the number of reconstruction iterations appears to narrow the PSFs of the beads. This effect is visualized in Fig. 3b, where MIPs of the mean PSF of all the point targets inside the test-object for 1, 10 and 100 PA reconstruction iterations are shown in the *xy* and *yz* plane. For 10 reconstruction iterations and 101 bowl rotational steps, FWHMs of 786 $\mu$m in *x*, 775 $\mu$m in *y* and 693 $\mu$m in *z* directions were estimated. The spatial resolution can be improved to 426 $\mu$m in *z* when a 100 reconstruction iterations are used in combination with 101 or more bowl rotational steps (Fig. 3c).

The performance of the system in imaging speed of sound (SOS) was evaluated with UST measurements on a well-characterized sound speed test object[24] that consists of multiple materials. Figure 3d shows two slices of the ground truth sound speed distribution of the test object in the imaging bowl. The corresponding slices of the SOS distributions (Fig. 3e), recovered from the UST mode measurements using a bent-ray inversion approach in 3D[28], show the composite structure of the test object and surrounding water. The 8 mm diameter filled channels inside the object are also visualized. The sound speeds across material interfaces are smeared due to the inherently limited spatial resolution achievable with a ray-based reconstruction. This is also seen in Figure 3f, which shows the SOS profiles along the dashed lines in Figs. 3d and 3e. The average absolute deviation of the SOS reconstruction from the ground truth on this profile is 11 m/s (or 0.7%).

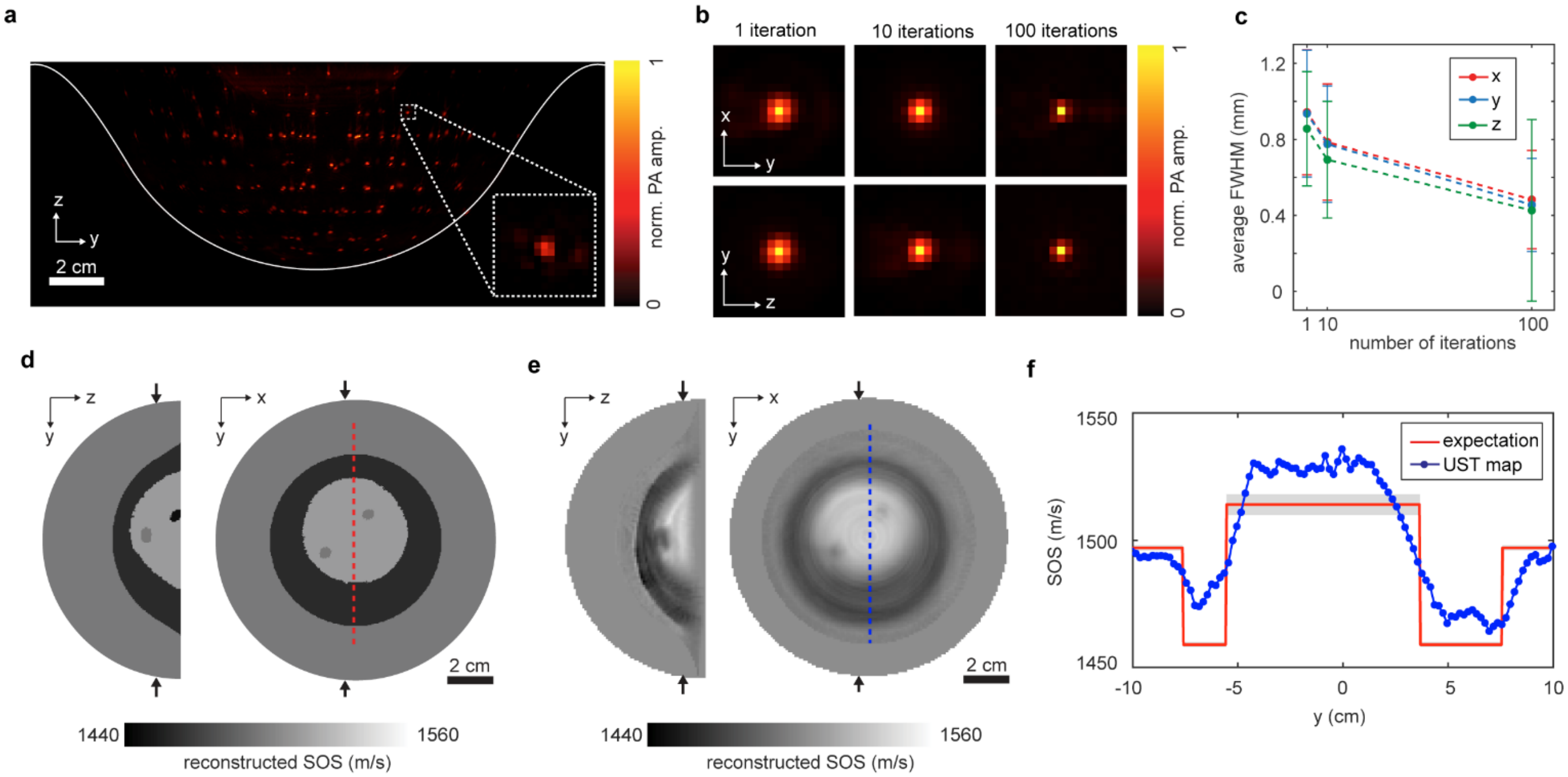

**Figure 3.** Photoacoustic (PA) and speed of sound (SOS) imaging performance on test objects and phantoms. **a,** A maximum intensity projection (MIP) from the PA measurement of the PSF test object. The inset shows a zoom-in on one of the beads showing the isotropic nature of the PSF. **b,** MIPs of the average PSF of all the reconstructed beads in the test-object for 1, 10 and 100 reconstruction iterations in the xy and yz plane. This shows that increasing the number of reconstruction iterations narrows the PSF. The spread and mean of Gaussians fitted to all PSFs in the test object are presented in **d,** Ground truth sound speed distribution in the SOS test object in *xy* and *xz* planes. The SOS values were measured on reference objects using transmission measurements in an independent characterization set-up. **e,** Slices of the reconstructed SOS map at the same locations as in **d**. **f,** The sound speed profiles along the dashed lines in **c** and **d**.

## In vivo imaging

We imaged both breasts each of eight healthy female volunteers; four exemplary breast images of four individuals are included, of which we discuss two in detail. PA images of all other imaged breasts can be found in Supplementary Fig. A4. The image acquisition times ranged from 3 to 6 minutes as different combinations of wavelengths, bowl rotations, and signal averages, were experimented with aiming to develop an optimal measurement protocol for future studies. Detailed information about the parameter setting can be found in Supplementary Table A2.

Volunteer 1 was a 64 year old woman (brassiere size 85A), who had best fit to breast cup 3. Volunteer 2 was 51 years old (brassiere size 90D) and breast cup 6; volunteer 3 was 55 years old (brassiere size 90D) and breast cup 8;volunteer 4 was 58 years old (brassiere size 80 C) with breast cup 5. All four volunteers had skin tones matching category 2 out of 5 on the Fitzpatrick scale[29] (Supplementary Tables A3 and A4).

Figure 4a shows depth color-coded anterior-posterior (AP), medio-lateral (ML) and cranial-caudal (CC) MIPs of the full-SOS compensated PA reconstruction at 800 nm of the right breast of volunteer 1. The images reveal complex vasculature throughout the well-centered breast with a high contrast and a wide and uniform field-of view. The superficial vessels (0-0.35 cm deep) located within the dashed box in the AP MIP, are shown in the zoomed in view in Figure 4b. The intensity profiles of three vessel cross-sections, fitted with a Gaussian, show that vessels with diameters of at least 0.78 mm and larger are depicted.

The full-SOS compensated 3D PA images for the three other volunteers are shown in Figure 4c, with wavelengths indicated in the images. For all cases, the nipple is visible at the surface (white arrow) and large vessels, presumably branches of the internal (asterisk) and lateral thoracic arteries (hollow arrowhead), show branching formations a few mm beneath the skin. Several deep vessels are also visible; the deepest vessels can be traced up to 34 mm, 48 mm, 43 mm and 42 mm depths for the four volunteers respectively. Several deep vessels are oriented perpendicular to the chest wall (see ML view of volunteer 3) and may be intercostal arteries that feed the deep breast parenchyma or they may originate from superficial veins that drain to the center of the breast[30]. Another observation is that two vessels run closely in parallel or revolve around each other at several locations (circle). These likely are the venous and arterial anatomies that parallel each other[31], especially for intercostal, axillary and internal thoracic vascular pathways.

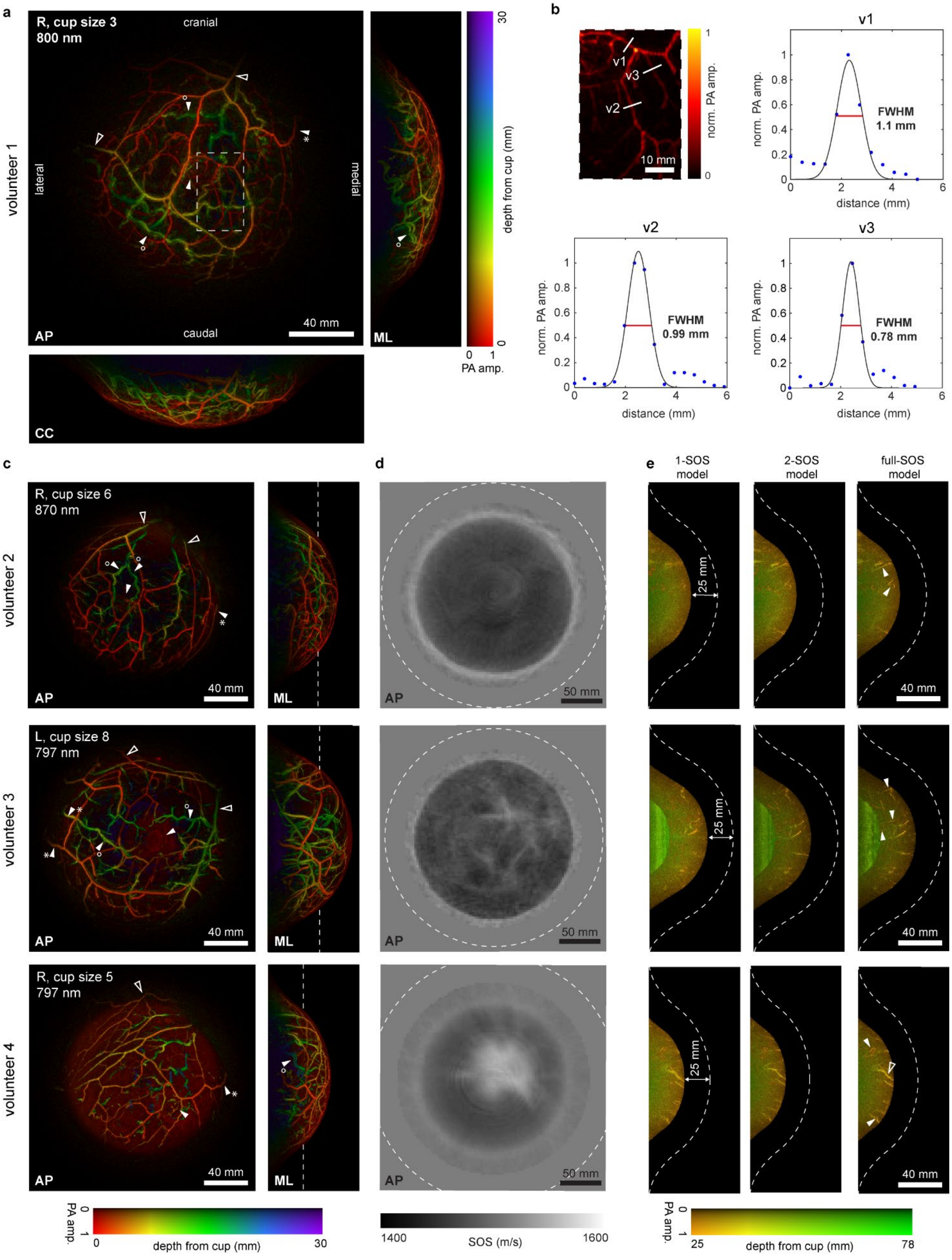

**Figure 4.** In vivo PA and SOS breast images made with thePAM3 system **a,** full-SOS compensated AP, ML, and CC PA MIP images of the right breast of healthy volunteer 1, demonstrating the system's capability for full breast imaging. **b,** A zoomed-in view of the superficial vessels (0-0.35 cm deep) in the ROI that is indicated by the white dashed box in a. Cross-sectional profiles of three vessels (v1, v2 and v3) are plotted and fitted with Gaussians to measure the vessel diameters. **c,** full-SOS compensated AP and ML PA MIPs of volunteers 2, 3 and 4 to further demonstrate the performance of the system. Locations of the nipple (white arrowhead), parallel vessel pairs (circle) and expected branches from the internal (asterisk) and lateral thoracic arteries (hollow arrowhead) can be observed in the images. The yellow dashed lines in the ML MIPs show the z-coordinate at which the AP slices of the SOS maps are presented in **d.** The dashed yellow circles in the SOS maps correspond to the contour of the bowl surface. **e**, ML MIPs of the deeper part of the breast for reconstructions using the 1-SOS, 2-SOS or full-SOS model reconstructions. The breast cup contour which corresponds to the surface of the breast is indicated by the white dashed lines. Solid arrowheads highlight vascular structures that are improved with the full-SOS compensated reconstruction, and hollow arrowheads point at structures that experience the opposite. All PA images in this figure are amplitude modulated.

Figure 4d shows slices of the SOS maps acquired from the UST measurements on volunteers 2, 3 and 4. Large-scale to mid-scale acoustic property heterogeneity is seen with sound speed variations in the range from 1370-1575 m/s, which match with values expected in the human breast[32,33]. The three volunteers show different sound speed distributions within the breast, which could indicate different breast parenchymal patterns[34,35]. The homogeneous and relatively low sound speed in volunteer 2 indicates a fatty breast. The inhomogeneous SOS distribution in volunteer 3 is likely a breast where fibrous or glandular tissue is interwoven with fat. The SOS distribution in volunteer 4 points to a centrally located region of dense fibroglandular tissue surrounded by a layer of fat tissue.

The effect of using our sophisticated SOS model in the PA inversion was investigated by comparing PA reconstructions from the full SOS map from the UST measurements (full-SOS) with a single SOS value (1-SOS) and two different SOS values outside and inside the breast (2-SOS). Figure 4e shows ML MIPs of volunteers 2 to 4. To examine deeper-lying vasculature, the first 25 mm were digitally peeled off in the reconstructed volumes (Figure 4e). These MIPs show that the 1-SOS model only results in groupings of higher intensity pixels, which appear to merge in the 2-SOS approach to form elongated structures rising above noise. The full-SOS compensation results in improvements in sharpness or increase in length (marked with solid arrowheads) of vessels, with appearance of a few new structures. There are however also a few structures that lose sharpness with the full-SOS approach compared to the 2-SOS approach (marked with hollow arrowheads). The green semi-circle in the deep breast of volunteer 3 is an artifact that only appears in measurements with cup sizes seven and eight. It originates from the PA signals generated on the surface of the imaging bowl that are back reflected from the breast surface and detected.

**Multi-wavelength in vivo imaging**

The left breast of volunteer 3 was imaged with a multi-wavelength measurement sequence (720, 755, 797, 833 and 860 nm) in a measurement time of 6 minutes. AP MIPs of the PA reconstructions with 720, 797 and 860 nm excitation (Fig. 5a) reveal the now familiar vascular architecture, but also melanin-rich structures at the nipple (white arrowhead), the skin and birthmarks on the skin (hollow arrowhead). The PA intensity of these latter structures decays with increasing wavelengths as expected from the known monotonically decreasing melanin absorption spectrum[36]. An interesting observation pertains to a structure located a few millimeters beneath the skin (enclosed in elliptical box). This structure is clearly visible in the 720 nm measurement but becomes indistinguishable at 860 nm. The distinctive form suggests that it is a blood vessel, and its diminishing visibility at longer wavelengths implies that it carries blood with a relatively lower oxygenation level. Figure 5b reveals the vascular structures at depths of 25 mm and beyond as a ML MIP for the same three wavelengths as in Figure 5a. The signal-to-noise ratio (SNR) of the vessels decreases with increasing wavelength. The radiant exposure is not same for all wavelengths but follows Fig. 2a. While a vessel in the 720 nm measurement (9 mJ $cm^{-2}$) was traced down to a 48 mm depth, the same vessel was traced down to 43 mm in the 797 nm measurement (7.5 mJ $cm^{-2}$) and down to 39 mm in the 860 nm measurement (6.7 mJ $cm^{-2}$). The decaying SNR with wavelength is also seen in Figure 5c, which plots maximum PA intensity as a function of depth encountered in 5 mm thick layers going radially inwards from the breast surface. This behavior can be attributed to the wavelength-dependent laser output energy with progressively lower energies at longer wavelengths, combined with the simultaneous increase in (coupling) water absorption at the longer wavelengths until its eventual domination in the near-infrared wavelength regime. The increase in intensity after 40 mm depth can be attributed to an enhanced sensitivity due to focusing effect of the hemispherical detection aperture towards the center of the bowl. The influence of noise and interference signals begins to dominate at a larger depth around 60 mm.

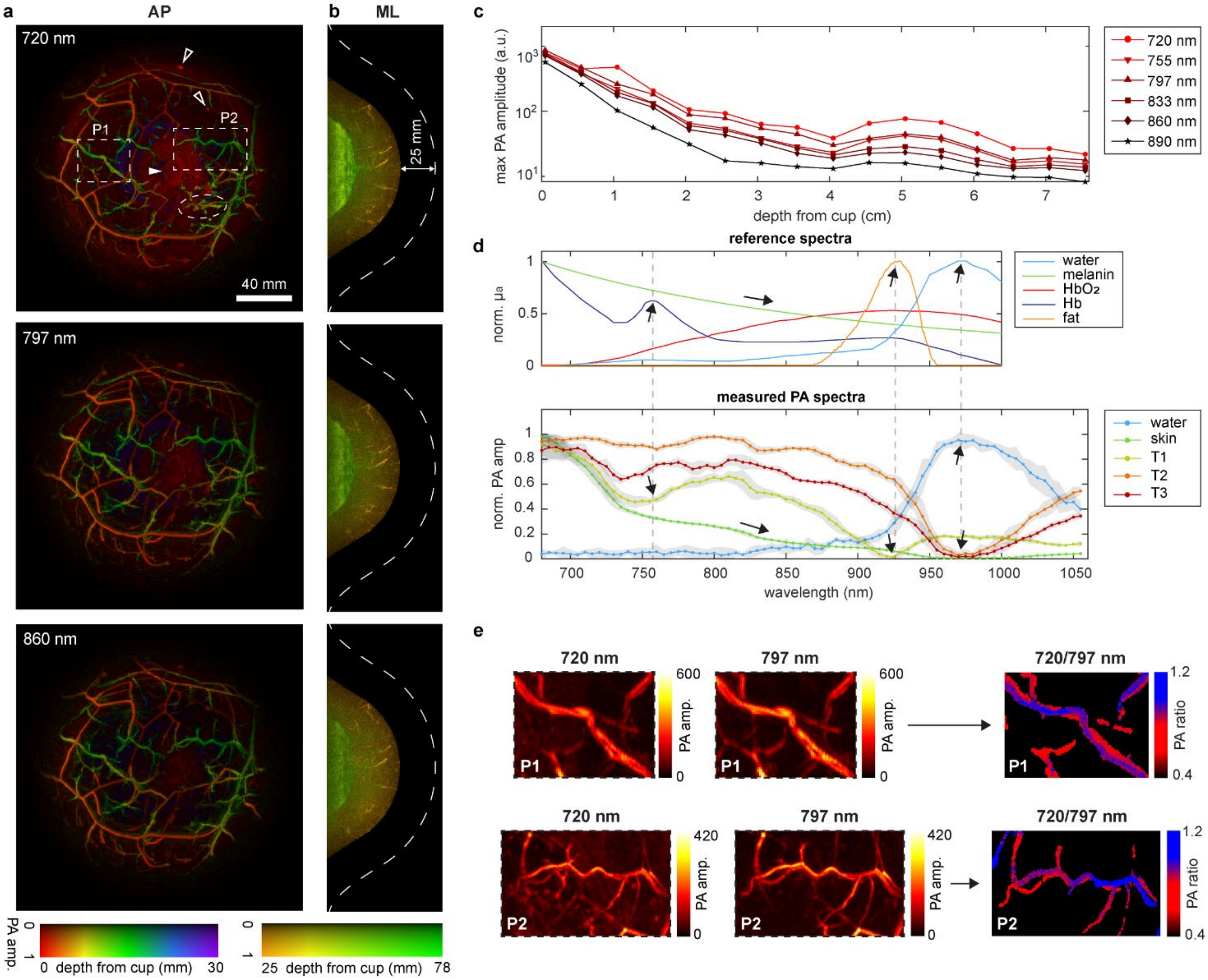

**Figure 5.** Multispectral in vivo imaging**. a,** Amplitude modulated AP PA MIPs of the left breast of a healthy breast (volunteer 3) for three out of the six wavelengths measured with. All images are normalized to their own maximum value. The solid arrowhead in the first image points to the nipple, the hollow arrowheads to birthmarks and the asterisk to a presumably vascular structure. All these indicated structures strongly decay in intensity with increasing wavelength. **b**, ML MIPs of the same PA reconstructions, but only revealing the structures at a depth of 25 mm and beyond. the maximum PA intensity as a function of depth from the cup surface for all six wavelengths. **c**, The maximum PA intensity encountered in 5 mm thick layers going radially inwards from the breast surface for the six different wavelengths imaged with. **d**, PA spectra corresponding to the five peaks in the time trace recorded by the lowest detector in the bowl during a measurement with the bowl static and the laser sweeping over the full wavelength range. **e** AP and CC MIPs of two ROIs containing parallel vessel pairs (P1 and P2). The locations of the pairs in the reconstructed volume are outlined in a. The 720/797 nm ratio images of the segmented vessel pairs show that one vessel contains more oxygen than the other.

Spectral signatures of tissue chromophores can also be observed in the recorded PA signals. The amplitudes of five peaks in the time trace recorded by the transducer at the lowest point in the bowl during a wavelength sweep measurement on the left breast of volunteer 3 are plotted as a function of wavelength in Fig. 5d. The five peaks were selected as follows: one was located at a time coordinate corresponding to the water in the imaging bowl, one at a coordinate corresponding to the skin, and the remaining peaks (T1, T2, T3) at later time coordinates belonging to tissues lying deeper in the breast. The time trace and the selected peaks can be found in the Supplementary Figure A5. The behaviour of the five peaks individually as function of wavelength demonstrate the influence of the spectra[36] of the prominent chromophores water, melanin, hemoglobin and fat (see Fig. 5d). While the absorption spectrum of water is directly recovered from the peak corresponding to water, the peaks at coordinates inside tissue show the integrated effect of the spectra of various chromophores encountered in the breast on the spherical surface centered on the element and at distance ($v$ x $t$). The effect seen is that of spectral coloring by various tissue chromophores[37,38], for example in the spectrum of T1, the local minimum at 755 nm corresponds to the local peak of Hb at the same wavelength, while the local minimum around 930 nm corresponds to the peak absorption in fat at the same wavelength.

Two parallel vessel pairs (P1 and P2) present in the breast show in the AP MIP within the ROIs outlined by the white dashed lines (Fig. 5a). Figure 5e presents the zoomed-in AP and CC views of these ROIs. Pixel-pixel ratio imaging was performed with images at 720 and 797 nm of these vessel pairs. These ratio images consistently show higher ratio values in the one vessel than in the other from which we can conclude that one is an artery and the other a vein.

## Discussion

We have developed the breast imaging system PAM3, that for the first time combines multi-wavelength PA tomography with UST. The system is a major technological improvement over the state-of-the-art due to the following advancements in critical hardware and software sub-systems: 1) the dual laser and optical parametric oscillator (OPO) units have a combined output with the highest intensity pulses in their class of laser, from 450 to 230 mJ/pulse between 680 to 1060 nm, 2) the light delivery approach using optical fiber bundles to provide 40 injection points distributed appropriately and with diverging optics, leads to homogeneous exposure with high fluence on the breast surface, 3) the 512 US transducers demonstrate a 123% fractional frequency reception bandwidth centered at 1 MHz; the large active area and carefully designed matching/backing layers realize an MDP of 0.3 Pa which is the lowest reported in literature for photoacoustic breast tomography, 4) the specially designed US pulser-receiver units provide 512-channel low-noise fully programmable electronic I/O modules, 5) the novel refraction-corrected 3D ray approach[28] demonstrates good quantitative accuracy for sound speed reconstruction from UST measurements, and 6) the novel PA image reconstruction method using an accelerated, iterative, first-order optimization procedure[27] can utilize the measured sound speed map in PA reconstruction, as well as take into account the spatial and temporal detector responses.

The system was tested and optimized on a specially developed suite of task-based test-objects[24] and a sophisticated PA-US breast phantom[25,26], and its capabilities and functionalities demonstrated on healthy human volunteers. The sound speed maps from the UST mode on human breasts revealed large-scale and medium-scale inhomogeneities with values and 3D distributions as can be expected in the breast. The volumetric sound speed data was used to correct perturbations to the wavefront in PA image reconstruction. The combination of this approach with the technical improvements mentioned above has enabled *in vivo* results that represent unprecedented imaging depth (as much as 48 mm), wide field of view and excellent contrast in the visualization of vascular morphology in the breast at multiple wavelengths.

In addition to improving PA reconstructions, the 3D sound speed images also provided insights into tissue-type distribution in the breast with the lower values suggesting the presence of adipose tissue and higher values the presence of fibroglandular tissue. Sound speed has diagnostic potential having been used in UST for breast density characterization and for identifying tumors based on speed of sound contrast[40,41]. Future technical research will pursue the development of full waveform inversion schemes for extracting sound speed with the potential for achieving higher resolution[42]. A pressing topic is the development of algorithms to extract the US reflectivity from the detector data. This will enable to extract, in addition to the optical absorption and sound speed contrasts already possible, an image based on also echogenicity contrast which is one of the mainstays of standard US imaging and is showing much promise in UST[40,41].

We demonstrated that the PAM3 system is capable of performing multi-wavelength *in vivo* studies requiring for a standard imaging protocol 5 minutes for a 5 wavelength 4 average tomographic image acquisition at 10 Hz. Future work will be in developing and advancing quantitative image reconstruction algorithms for extracting measures of blood oxygen saturation in tissue. The ability to estimate this parameter from tumor vasculature is thought to be key to imaging cancer reliably[39]. Further, multi-wavelength measurements can also allow the estimation of concentrations of other tissue chromophores in the breast such as lipid and water, which are thought to be different between tumor and healthy tissues[44]. The first step towards quantitation this has been taken in our approach to solve the acoustic part of the inverse problem accurately with the use of an heterogeneous distribution of speed of sound in the breast acquired from the UST mode.

While the potential of PA and US tomography in the context of breast cancer has been shown[4,40,41,43-51], it is important to acknowledge that there are still gaps in understanding and interpreting PA breast images. Further, the precise roles for photoacoustic imaging in the breast cancer imaging paradigm, whether in screening, diagnosis, treatment planning, neoadjuvant chemotherapy monitoring or follow-up are not fully known. Whether the method is suitable for specific patient populations such as women with dense breasts is also not known. With the new imager, which has reached the required level of technical sophistication enabling high quality spectral PA and sound speed imaging, we believe the time is ripe to embark on focused clinical studies to address the lacunae in our understanding and the open questions in our field.

## Methods

### Additional system information

The document 'Supplementary materials' carries more information about the system including the laser, light delivery, system control and data storage.

### Photoacoustic inversion

The PA images are estimated for each excitation wavelength from the measured PA time traces using an iterative, full-wave, model-based image reconstruction method which is expected to lead to quantitatively accurate solutions[27]. The wave propagation part can use three different SOS models, reflecting an increasing degree of accuracy: Using the SOS of water at 25°C (which corresponds to the coupling medium in the bowl) everywhere in the volume will be called "1-SOS model" but does not account for SOS heterogeneities. Using a binary mask of the breast cup to assign a different SOS value inside the breast will be called "2-SOS model". The value inside the breast can be manually or automatically tuned to obtain sharp blood vessels[52]. Using the SOS map obtained from US transmission measurements to compensate for SOS heterogeneities will be called “full-SOS model”. The numerical implementation of the wave propagation is based on the k-Wave toolbox[53] but uses a tailor-made CUDA implementation that also includes part of the US transducer modeling for memory efficiency. The US sensing performance of the transducers is modelled to account for their directional response and the empirically measured temporal impulse responses and relative sensitivities of all the elements. (See Sec. US transducer characterization for the corresponding calibration measurements.) All PA images in this manuscript were reconstructed on an isotropic computational grid of 0.4 mm using a TITAN RTX GPU by NVIDIA. A single wave simulation on this grid takes 10 minutes to run and supports frequencies up to 1.87 MHz, which means that all the relevant detector bandwidth is covered.

### Quantitative sound speed imaging

For quantitative SOS imaging, a ray-based approach using a two-point ray tracing scheme (employing a shooting method), was used to calculate the SOS from the time-of-flights (TOF) of acoustic waves between emitter-detector pairs[28]. The TOFs are computed using a first-arrival algorithm, based on the Akaike-Information Criterion[54]. The SOS map was reconstructed by minimizing the norm of the discrepancy between the experimental and modelled TOFs through a succession of linearisations[28]. A reference UST measurement of the water-filled bowl was used to increase the quantitative accuracy by reducing uncertainties in the TOF-picking, detector positions or detector frequency responses. Two approaches to the minimisation were used. In ‘absolute’ imaging, the recovered unknown parameter was the SOS map within the region of interest and the linearisations were made about the most recent estimate. In ‘difference’ imaging, the unknown parameter was the difference of the sound speeds between the breast and the water-only reference, and the linearisation point was chosen to be between the known water reference sound speed and the last estimate of the sound speed map. Note that both approaches reconstruct the sound speed images from the difference of the time-of-flights for the object-in-water and only-water data, and where they differ is only on the approach taken for choosing the point of linearization. The absolute approach recovered sharper interfaces in sound speed distributions, and so was used for visualizing the SOS maps (Fig 2h and 4d). The relative approach led to SOS maps that better correct for aberrations and was therefore used in the PA inversion to obtain sharper PA images. The SOS maps were reconstructed on a 2 mm grid spacing in few hours on a single CPU.

**US transducer characterization**

The characterization measurements were performed in an external set-up on all 512 US elements before the elements were installed in the imaging bowl. Each US element was connected to a 40 dB low noise amplifier. The impedance of amplifier and the cable lengths used in these characterization measurements were similar as in the PAM3 imager.

Noise measurements were performed with a spectrum analyzer (Agilent 8594E) in the 0.2-2 MHz frequency range. The US element was placed in a Faraday cage in order to minimize the environmental noise. The US element electrical impedance was measured with a network analyzer (Agilent 4294) in the same frequency range and it was checked that the noise measurement results fit well with a theoretical calculation based on the transducer impedance (thermal noise) and the amplifier noise.

The sensitivity of the US elements was measured with a substitution method. The transmitter was a large bandwidth transducer. The central frequency and bandwidth of the reference transducer are: $f_c$ = 1 MHz, $f_u$ = 0.592 MHz and $f_l$ = 1.50 MHz -6dB for a BW of nearly 1 MHz. ($f_c$ is centre frequency, $f_l$ is lower frequency and $f_u$ is upper frequency)

The reference transducer was excited with a chirp signal (0.1-2 MHz bandwidth) with a generator (Keysight 33600A, USA). The incident pressure produced by the transmitter was measured with a PVDF hydrophone (Onda HGL 0400). This hydrophone was then replaced by the US element connected to the 40 dB amplifier and the generated voltage ($V_{out}$) on the US element was recorded. The receive transfer function (RTF) (V/Pa) was calculated from the incident pressure $P_{in}$ and the measured voltage ($V_{out}$) using RTF = 20 log ($V_{out}/P_{in}$).

The frequency responses of all transducers were measured after installing the detectors in the system. In this case, all elements were characterized simultaneously by means of a PA measurement on a specially designed black phenolic ball[24] at 755 nm with 121 bowl positions and 16 measurements per position (averages). All the time-traces measured with one detector (121x16 signals) were shifted in time to have all peaks overlapping. These shifted time traces were averaged and the frequency response was calculated for each detector by Fourier transforming this averaged signal.

**Light distribution on the cup surface**

A set of PA measurements was performed on black variants of the eight breast-supporting cups[24] to visualize the light distribution on the surface of each. The black cups were produced by vacuum forming a 1 +/- 0.1 mm thick black PVC-U sheet over the same moulds that are normally used to produce the transparent breast-supporting cups[22]. The cups were taped to the same metal rings that are normally taped to the disposable transparent cups. With this ring, the cups were one-by-one installed in the imaging aperture and a PA measurement at 755 nm using 201 bowl positions and 10 PA averages was performed on each of the cups. The PA images were reconstructed on a 0.4 mm grid using 10 reconstruction iterations.

## Spatial resolution measurements

A test-object was developed to measure the PA spatial resolution over the entire imaging volume in the three imaging dimensions. The object contains 277 sub-resolution India ink coated stainless steel microspheres (Cospheric LLC, United States) that have a 310-360 µm diameter which are distributed throughout a cup 8 volume and held in a polyvinyl plastisol (PVCP) matrix[24]. The PVCP (LureFlex firm, LureFactors, UK) has a 1403±3 m/s sound speed at room temperature and at 1 MHz and is optically and acoustically semi-transparent. A highly sampled PA measurement was performed on this object, consisting of 401 bowl rotations and 30 averages. Reconstructions were made with the full dataset but also with down sampled sets of the data to investigate the effect of the number of bowl positions, averages and reconstruction iterations on the spatial resolution.

## Validation of SOS imaging

A test object containing materials with known sound speeds was imaged to test the ability of the UST modality in calculating SOS maps.[24]. An MRI scan was made and segmented to have a 3D model of the object (see Fig. 3f). UST measurements using the same transducers and electronics as are used in the imager were performed on the materials to characterize their sound speeds. Two spare transducers and a spare I/O board were used to make one transducer send pulses with a 1 MHz center frequency and to have the other transducer in reception mode. Time-of-flights were computed from the recorded time-traces using the same first-arrival-picking algorithm as in our UST reconstruction codes.

The object was imaged in the imager with 101 bowl positions and with 513 US shots per bowl position, which is the maximum amount of US shots supported by the DAQ. The SOS map was reconstructed using the quantitative SOS reconstruction approach described above.

## Volunteer recruitment and measurements

The Medical Research Ethics Committees United (MEC-U), acknowledged by the Dutch Central Committee for research involving human subjects, reviewed and approved the study. An independent party (DARE!! Measurements, The Netherlands) performed product safety tests on the PAM 3 system, which the local hospital used to approve the release of the imager for measurements on human subjects. The study was registered in the Netherlands National Trial Register under the identification number NL7992.

In total eight healthy volunteers, aged between 23 and 64 years, were recruited for the study. The most important exclusion criteria included pregnancy or breastfeeding, a history of a breast disease and the presence of a tattoo or piercing in the breast area. Each volunteer was informed about the study and informed consent was obtained in all cases. For each volunteer, a short questionnaire was used to classify the skin tone on the Fitzpatrick scale[29]. The skin types and the questionnaire are included in Supplementary materials (Table A3 and A4)

Multiple measurements were performed on each volunteer, within a maximum measurement time of 20 minutes per breast. A standard PA-US measurement protocol consisting of 101 bowl positions takes 5 minutes for 5 wavelengths with 4 averages, but since measurement sequences were not the same for all volunteers, individual measurement times are not the same. The number and choice of wavelengths, number of bowl rotation steps and number of PA averages are examples of settings in the sequence that were varied in the study to investigate how these parameters influence the results, from which the settings for optimal imaging quality were extracted.

For every measurement, the appropriate cup size was selected and the cup was installed in the imaging aperture. The breast to be imaged was disinfected and the volunteer positioned prone on the bed with her breast in the cup. The volunteer wore laser goggles as protection against possible laser light escaping from the bowl. The arm contralateral to the breast being imaged was placed above the head and the contralateral leg was bent outwards to incline the torso. This specific way of positioning increased the volume of breast tissue available in the imaging bowl. Pillows for the head, the hips and the ankles were available and were used in case the volunteer experienced discomfort. Once the volunteer was comfortable, the operator at the control PC made sure that the water level in the bowl was topped off and then started the measurement. The described protocol was repeated for the contralateral breast. After the measurements, both breasts were visually inspected for birthmarks, scars and other marks on the skin and their locations were noted down by the operator.

**Image analysis and visualization**

**Spatial resolution calculation**

Additional information can be found in document ‘Supplementary Materials’.

**Ratio imaging**

For ratio imaging, the reconstructions for two wavelengths (720 nm and 797 nm) were normalized by the laser output energy at the respective wavelengths. For both reconstructions, the same ROI was selected and the vascular structures of interest were segmented. Additionally, both ROIs were normalized to the maximum value of the isosbestic point (797 nm) reconstruction. The ratio image was then calculated by dividing the 720 nm ROI by the 797 nm ROI.

**Visualization of in-vivo images**

No motion correction was employed since a first investigation revealed no discernable breathing artefacts in the images. For image visualization, the reconstructions were first Frangi-filtered to enhance the contrast of the vessels[55,56]. Then adaptive intensity modulation (AIM)[56] was applied to homogenize the pixel intensities by suppressing the strong superficial pixels and enhancing the deep low intensity structures. AIM was implemented by dividing the Frangi-filtered image by a standard deviation map that was calculated by sliding a 10x10x10 pixel window over the image in all dimensions with a 3 voxel step size[56]. The final image was a weighted summation of the original image and the intensity modulated image[5,56]. Weighting factors that gave the visually most compelling images were used and yielded 0.65 and 0.35 respectively.

For depth-encoding the images, we made use of the well-defined breast position and cup contour[22]. This allowed the shortest distance to the breast surface to be calculated for each voxel within the cup volume. A 2D colormap was prepared with color hues on one axis representing the depth from the skin and the color value on the second axis representing the PA intensity.

## Data availability statement.

The research data that underlie this publication will be shared on acceptance of the manuscript, via a University of Twente server together with relevant meta-data, and in line with FAIR (Findability, Accessibility, Interoperability, and Reusability) principles.

## Code availability statement

The computational code for reconstructing sound speed images and photoacoustic images as reported in this publication will be made available to readers on acceptance of the manuscript.

## Acknowledgements

The PAM3 imaging system was developed in the PAMMOTH project, made possible by the Horizon 2020's research and innovation program, H2020 ICT 2016-2017, under grant agreement No 732411 which is an initiative of the Photonics Public Private Partnership. Author BC was also supported by grant EP/T014369/1 from the UK's Engineering & Physical Sciences Research Council. Author JJ was also supported by the Ministry of Education, Youth and Sports of the Czech Republic through the e-INFRA CZ (ID:90140) project under the program Projects of Large Research, Development, and Innovations Infrastructures, which provided access to CESNET storage facilities. The authors thank Bradley Treeby, Roeland Huijink, David Thompson, Michael Jaeger and Martin Frenz for useful inputs in the design phase of the system; all students who contributed during their Bachelor or Master projects; TP21 for assistance in managing the project; the Medisch Spectrum Twente Mammapoli and breast radiologists for enabling and assisting with the *in-vivo* measurements. Cindy Lammertink and Job van der Palen provided valuable inputs to improve the Medical Research Ethics study protocols. The PAMMOTH advisory committee (Prof. R. Pijnappel, Prof. M. Broeders, Prof. J. Wesseling and Ms. E. Verschuur) are thanked for their useful feedback.

## Author contributions

All authors contributed during the system design phase; FL, AJ and MD performed simulations during this phase. EC, MW, FB and TM developed the US detectors, and TK and AM the laser and fiber bundle design. LA developed the DAQ system, RPPM the system mechanics and patient-user interface. MN and LA developed the system control software. FL and AJ developed reconstruction codes, under guidance of BC. JB, GB, supervised by JJ, developed codes to pre-process the recorded data before the reconstruction. MD designed the human study protocol with inputs from LFGO, SA, JV and SM. MD wrote the medical device dossier with inputs from LA, RPPM, MN, and did laser safety calculations reviewed by SA. SCK prepared risk analysis and other documents for the ethical review committee under the supervision of SA with inputs and contributions from MN, MD, RPPM. MD and SCK conducted test-object and phantom measurements, and healthy volunteer measurements. FL and AJ reconstructed images. MD and SCK analyzed the images with inputs from FL. MD wrote the paper together with SM, with inputs from FL, AJ on the image reconstruction parts. FL, AJ, BC, LFGO and SM reviewed the paper. SM, BC, JJ, WMK, AM, MF, TM, JV and AM acquired financial support, and SM directed and supervised the project.

**Competing interests** None

## References

1. Wang, L. V., & Hu, S. (2012). Photoacoustic tomography: in vivo imaging from organelles to organs. science, 335(6075), 1458-1462.

2. Ntziachristos, V., & Razansky, D. (2010). Molecular imaging by means of multispectral optoacoustic tomography (MSOT). Chemical reviews, 110(5), 2783-2794.

3. Zhou, Y., Yao, J., & Wang, L. V. (2016). Tutorial on photoacoustic tomography. Journal of biomedical optics, 21(6), 061007.

4. Lin, L., & Wang, L. V. (2022). The emerging role of photoacoustic imaging in clinical oncology. Nature Reviews Clinical Oncology, 19(6), 365-384.

5. Lin, L., Hu, P., Tong, X., Na, S., Cao, R., Yuan, X., ... & Wang, L. V. (2021). High-speed three-dimensional photoacoustic computed tomography for preclinical research and clinical translation. Nature communications, *12(1), 1-10.*

6. Karlas, A., Pleitez, M. A., Aguirre, J., & Ntziachristos, V. (2021). Optoacoustic imaging in endocrinology and metabolism. Nature Reviews Endocrinology, 17(6), 323-335.

7. Na, S., Russin, J. J., Lin, L., Yuan, X., Hu, P., Jann, K. B., Yan, L., Maslow, K., Shi, J., Wang, D. J., Liu, C. Y. & Wang, L. V. (2022). Massively parallel functional photoacoustic computed tomography of the human brain. Nature Biomedical Engineering, 6(5), 584-592.

8. Haedicke, K., Agemy, L., Omar, M., Berezhnoi, A., Roberts, S., Longo-Machado, C., Skubal, M., Nagar, K., Hsu, H.-T, Kim, K., Reiner, T., Coleman, J., Ntziachristos, V., Scherz, A. & Grimm, J. (2020). High-resolution optoacoustic imaging of tissue responses to vascular-targeted therapies. Nature biomedical engineering, 4(3), 286-297.

9. Sung, H., Ferlay, J., Siegel, R. L., Laversanne, M., Soerjomataram, I., Jemal, A., & Bray, F. (2021). Global cancer statistics 2020: GLOBOCAN estimates of incidence and mortality worldwide for 36 cancers in 185 countries. CA: a cancer journal for clinicians, 71(3), 209-249.

10. Britton, P., Duffy, S. W., Sinnatamby, R., Wallis, M. G., Barter, S., Gaskarth, M., O'Neill, Caldas, C. Brenton, J. D., Forouhi, P. & Wishart, G. C. (2009). One-stop diagnostic breast clinics: how often are breast cancers missed?. British journal of cancer, 100(12), 1873-1878.

11. Nass, S. J., Henderson, I. C., & Lashof, J. C. (Eds.). (2001). Mammography and beyond: developing technologies for the early detection of breast cancer (pp. 1713-1722). Washington, DC: National Academy Press.

12. Pediconi, F., Catalano, C., Roselli, A., Dominelli, V., Cagioli, S., Karatasiou, A., Pronio, A., Kirchin, M. A. & Passariello, R. (2009). The challenge of imaging dense breast parenchyma: is magnetic resonance mammography the technique of choice? A comparative study with x-ray mammography and whole-breast ultrasound. Investigative radiology, 44(7), 412-421.

13. Hooley, R. J., Andrejeva, L., & Scoutt, L. M. (2011). Breast cancer screening and problem solving using mammography, ultrasound, and magnetic resonance imaging. Ultrasound quarterly, 27(1), 23-47.

14. Peters, N. H., Borel Rinkes, I. H., Zuithoff, N. P., Mali, W. P., Moons, K. G., & Peeters, P. H. (2008). Meta-analysis of MR imaging in the diagnosis of breast lesions. Radiology, 246(1), 116-124.

15. Manohar, S., & Dantuma, M. (2019). Current and future trends in photoacoustic breast imaging. Photoacoustics, 16, 100134.

16. Xu, M., & Wang, L. V. (2003). Analytic explanation of spatial resolution related to bandwidth and detector aperture size in thermoacoustic or photoacoustic reconstruction. Physical Review E, 67(5), 056605.

17. Tian, C., Zhang, C., Zhang, H., Xie, D., & Jin, Y. (2021). Spatial resolution in photoacoustic computed tomography. Reports on Progress in Physics, 84(3), 036701.

18. Lin, L., Hu, P., Shi, J., Appleton, C. M., Maslov, K., Li, L., Zhang, R. & Wang, L. V. (2018). Single-breath-hold photoacoustic computed tomography of the breast. *Nature communications*, *9*(1), 1-9.

19. Schoustra, S. M., Piras, D., Huijink, R., Op't Root, T. J., Alink, L., Kobold, W. M. F., Steenbergen, W. & Manohar, S. (2019). Twente Photoacoustic Mammoscope 2: system overview and three-dimensional vascular network images in healthy breasts. Journal of biomedical optics, 24(12), 121909.

20. Oraevsky, A., Su, R., Nguyen, H., Moore, J., Lou, Y., Bhadra, S., Forte, L., Anastasio, A. & Yang, W. (2018, April). Full-view 3D imaging system for functional and anatomical screening of the breast. In Photons Plus Ultrasound: Imaging and Sensing 2018 (Vol. 10494, pp. 217-226). SPIE.

21. Toi, M., Asao, Y., Matsumoto, Y., Sekiguchi, H., Yoshikawa, A., Takada, M., Kataoka, M., Endo, T., Kawaguchi-Sakita, N., Kawashima, M., Fakhrejahani, E., Kanao, S., Yamaga, I., Nakayama, Y., Tokiwa, M., Torii, M., Yagi, T., Sakurai, T., Togashi, K. & Shiina, T. (2017). Visualization of tumor-related blood vessels in human breast by photoacoustic imaging system with a hemispherical detector array. Scientific reports, 7(1), 1-11.

22. Schoustra, S. M., Op't Root, T. J., van Meerdervoort, R. P. P., Alink, L., Steenbergen, W., & Manohar, S. (2021). Pendant breast immobilization and positioning in photoacoustic tomographic imaging. Photoacoustics, 21, 100238.

23. Xia, W., Piras, D., Van Hespen, J. C., Van Veldhoven, S., Prins, C., Van Leeuwen, T. G., Steenbergen, W. & Manohar, S. (2013). An optimized ultrasound detector for photoacoustic breast tomography. Medical physics, 40(3), 032901.

24. Dantuma, M. M., Kruitwagen, S. C., Weggemans, M. J., Op't Root, T. J., & Manohar, S. S. (2021). Suite of 3D test objects for performance assessment of hybrid photoacoustic-ultrasound breast imaging systems. Journal of biomedical optics, 27(7), 074709.

25. Dantuma, M., van Dommelen, R., & Manohar, S. (2019). Semi-anthropomorphic photoacoustic breast phantom. Biomedical optics express, 10(11), 5921-5939.

26. Dantuma, M., Kruitwagen, S., Julia, J. O., van Meerdervoort, R. P. P., & Manohar, S. (2021). Tunable blood oxygenation in the vascular anatomy of a semi-anthropomorphic photoacoustic breast phantom. Journal of biomedical optics, 26(3), 036003.

27. Arridge, S., Beard, P., Betcke, M., Cox, B., Huynh, N., Lucka, F., Ogunlade, O. & Zhang, E. (2016). Accelerated high-resolution photoacoustic tomography via compressed sensing. Physics in Medicine & Biology, 61(24), 8908.

28. Javaherian, A., Lucka, F., & Cox, B. T. (2020). Refraction-corrected ray-based inversion for three-dimensional ultrasound tomography of the breast. Inverse Problems, 36(12), 125010.

29. Fitzpatrick, T. B. (1988). The validity and practicality of sun-reactive skin types I through VI. Archives of dermatology, 124(6), 869-871.

30. Jesinger, R. A. (2014). Breast anatomy for the interventionalist. Techniques in vascular and interventional radiology, 17(1), 3-9.

31. Vilagran, M., Del Riego, J., Palaña, P., Sentís, M., Planas, J., Oliva, J. C., & Gómez, V. (2019). Parallel vessels as a predictor of benignity in solid breast masses. The Breast Journal, 25(2), 257-261.

32. Szabo, T. L. (2004). Diagnostic ultrasound imaging: inside out. Academic press.

33. Duck, F. A. (2013). Physical properties of tissues: a comprehensive reference book. Academic press.

34. Pandya, S., & Moore, R. G. (2011). Breast development and anatomy. Clinical obstetrics and gynecology, 54(1), 91-95.

35. Wolve, J.N. Breast paranchymal patterns and their changes with age. Radiology, 121(3), 545-552.

36. Prahl, S. (1999). Optical absorption of hemoglobin. http://omlc. ogi. edu/spectra/hemoglobin.

37. Cox, B. T., Laufer, J. G., Beard, P. C., & Arridge, S. R. (2012). Quantitative spectroscopic photoacoustic imaging: a review. Journal of biomedical optics, 17(6), 061202.

38. Hochuli, R., An, L., Beard, P. C., & Cox, B. T. (2019). Estimating blood oxygenation from photoacoustic images: can a simple linear spectroscopic inversion ever work?. Journal of Biomedical Optics, 24(12), 121914.

39. Boas, D. A., & Franceschini, M. A. (2011). Haemoglobin oxygen saturation as a biomarker: the problem and a solution. Philosophical Transactions of the Royal Society A: Mathematical, Physical and Engineering Sciences, 369(1955), 4407-4424.

40. Gemmeke, H., Hopp, T., Zapf, M., Kaiser, C. and Ruiter, N.V., 2017. 3D ultrasound computer tomography: Hardware setup, reconstruction methods and first clinical results. Nuclear Instruments and Methods in Physics Research Section A: Accelerators, Spectrometers, Detectors and Associated Equipment, 873, pp.59-65.

41. Duric, N., Littrup, P., Roy, O., Li, C., Schmidt, S., Cheng, X. and Janer, R., 2014. Clinical breast imaging with ultrasound tomography: A description of the SoftVue system. The Journal of the Acoustical Society of America, 135(4), pp.2155-2155.

42. Lucka, F., Pérez-Liva, M., Treeby, B. E., & Cox, B. T. (2021). High resolution 3D ultrasonic breast imaging by time-domain full waveform inversion. Inverse Problems, 38(2), 025008.

43. Heijblom, M., Steenbergen, W., & Manohar, S. (2015). Clinical photoacoustic breast imaging: the Twente experience. IEEE pulse, 6(3), 42-46.

44. Diot, G., Metz, S., Noske, A., Liapis, E., Schroeder, B., Ovsepian, S. V., Meier, R., Rummeny, E. & Ntziachristos, V. (2017). Multispectral Optoacoustic Tomography (MSOT) of Human Breast Cancer. Clinical Cancer Research, 23(22), 6912-6922.

45. Becker, A., Masthoff, M., Claussen, J., Ford, S. J., Roll, W., Burg, M., Barth, P. J., Heindel, W., Schäfers, M., Eisenblätter, M. & Wildgruber, M. (2018). Multispectral optoacoustic tomography of the human breast: characterisation of healthy tissue and malignant lesions using a hybrid ultrasound-optoacoustic approach. European radiology, 28(2), 602-609.

46. Menezes, G. L., Pijnappel, R. M., Meeuwis, C., Bisschops, R., Veltman, J., Lavin, P. T., van de Vijver, M. J. & Mann, R. M. (2018). Downgrading of breast masses suspicious for cancer by using optoacoustic breast imaging. Radiology, 288(2), 355-365.

47. Neuschler, E. I., Butler, R., Young, C. A., Barke, L. D., Bertrand, M. L., Böhm-Vélez, M., Destounis, S., Donlan, P., Grobmyer, S. R., Katzen, J., Kist, K. A., Lavin, P. T., Makariou, E. V., Parris, T. M., Schilling, K. J., Tucker, F. L. & Dogan, 47. B. E. (2018). A pivotal study of optoacoustic imaging to diagnose benign and malignant breast masses: a new evaluation tool for radiologists. Radiology, 287(2), 398-412.

48. Oraevsky, A. A., Clingman, B., Zalev, J., Stavros, A. T., Yang, W. T., & Parikh, J. R. (2018). Clinical optoacoustic imaging combined with ultrasound for coregistered functional and anatomical mapping of breast tumors. Photoacoustics, 12, 30-45.

49. Kukačka, J., Metz, S., Dehner, C., Muckenhuber, A., Paul-Yuan, K., Karlas, A., Fallenberg, E. M., Rummeny, E., Jüstel, D. & Ntziachristos, V. (2022). Image processing improvements afford second-generation handheld optoacoustic imaging of breast cancer patients. Photoacoustics, 26, 100343.

50. Abeyakoon, O., Woitek, R., Wallis, M. G., Moyle, P. L., Morscher, S., Dahlhaus, N., Ford, S. J., Burton, N. C., Manavaki, R. Mendichovszky, I. A., Joseph, J., Quiros-Gonzalez, I. Bohndiek. S. E. & Gilbert, F. J. (2022). An optoacoustic imaging feature set to characterise blood vessels surrounding benign and malignant breast lesions. Photoacoustics, 27, 100383.

51. Schoustra, S. M., De Santi, B., op't Root, T. J., Klazen, C. A., van der Schaaf, M., Veltman, J., .Steenbergen, W. & Manohar, S. (2023). Imaging breast malignancies with the Twente Photoacoustic Mammoscope 2. Plos one, 18(3), e0281434.

52. Treeby, B. E., Varslot, T. K., Zhang, E. Z., Laufer, J. G., & Beard, P. C. (2011). Automatic sound speed selection in photoacoustic image reconstruction using an autofocus approach. Journal of biomedical Optics, 16(9), 090501-090501.

53. Treeby, B. E., & Cox, B. T. (2010). k-Wave: MATLAB toolbox for the simulation and reconstruction of photoacoustic wave fields. Journal of biomedical optics, 15(2), 021314.

54. Li, C., Huang, L., Duric, N., Zhang, H., & Rowe, C. (2009). An improved automatic time-of-flight picker for medical ultrasound tomography. Ultrasonics, 49(1), 61-72.

55. Frangi, A. F., Niessen, W. J., Vincken, K. L., & Viergever, M. A. (1998, October). Multiscale vessel enhancement filtering. In International conference on medical image computing and computer-assisted intervention (pp. 130-137). Springer, Berlin, Heidelberg.

56. Lin, L., Tong, X., Hu, P., Invernizzi, M., Lai, L., & Wang, L. V. (2021). Photoacoustic computed tomography of breast cancer in response to neoadjuvant chemotherapy. Advanced Science, 8(7), 2003396.

*********

# Supplementary material to: Fully three-dimensional sound speed-corrected multi-wavelength photoacoustic breast tomography

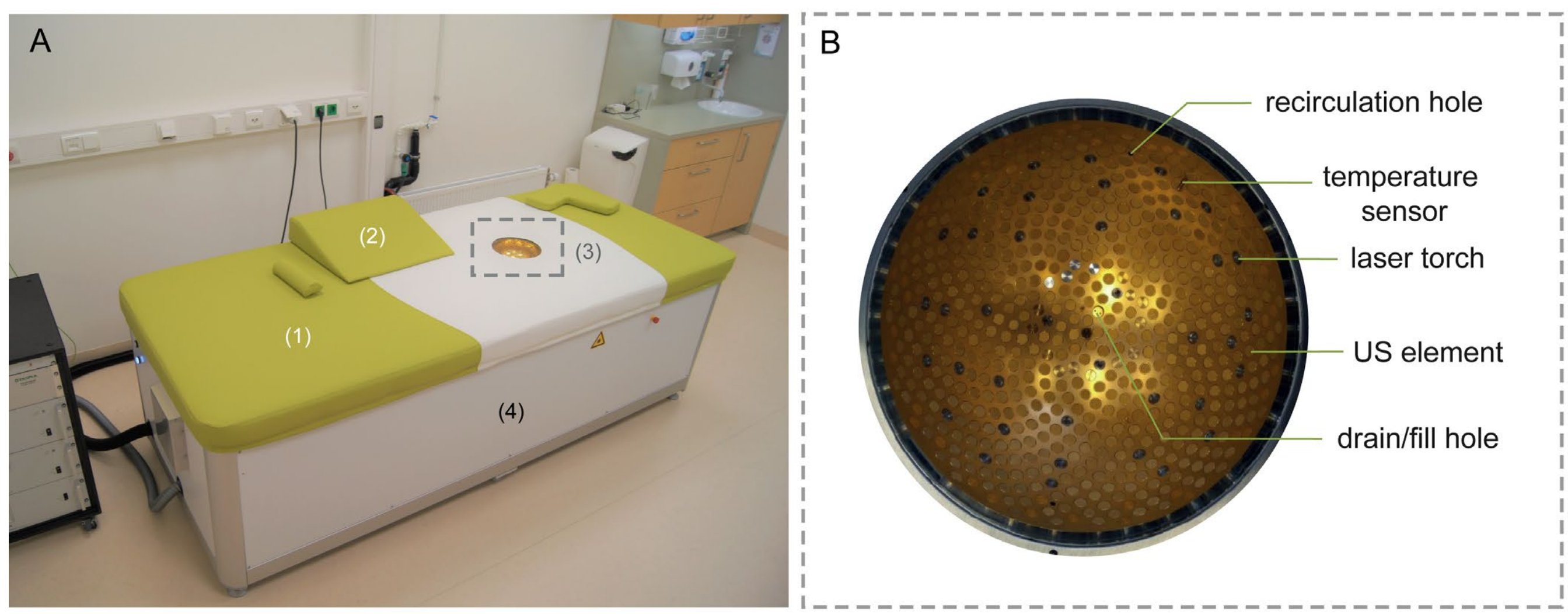

**Figure A1.** (A) Photograph of the PAM3 system with the bed (1) on which a woman can lie in prone position. Pillows (2) can be used to position the subject comfortably, and so as to have as much breast tissue available inside the imaging bowl (3). Most of the hardware is hidden behind the panels (4). (B) Photograph of top-view of the imaging bowl, showing the inserts in the spiral pattern.

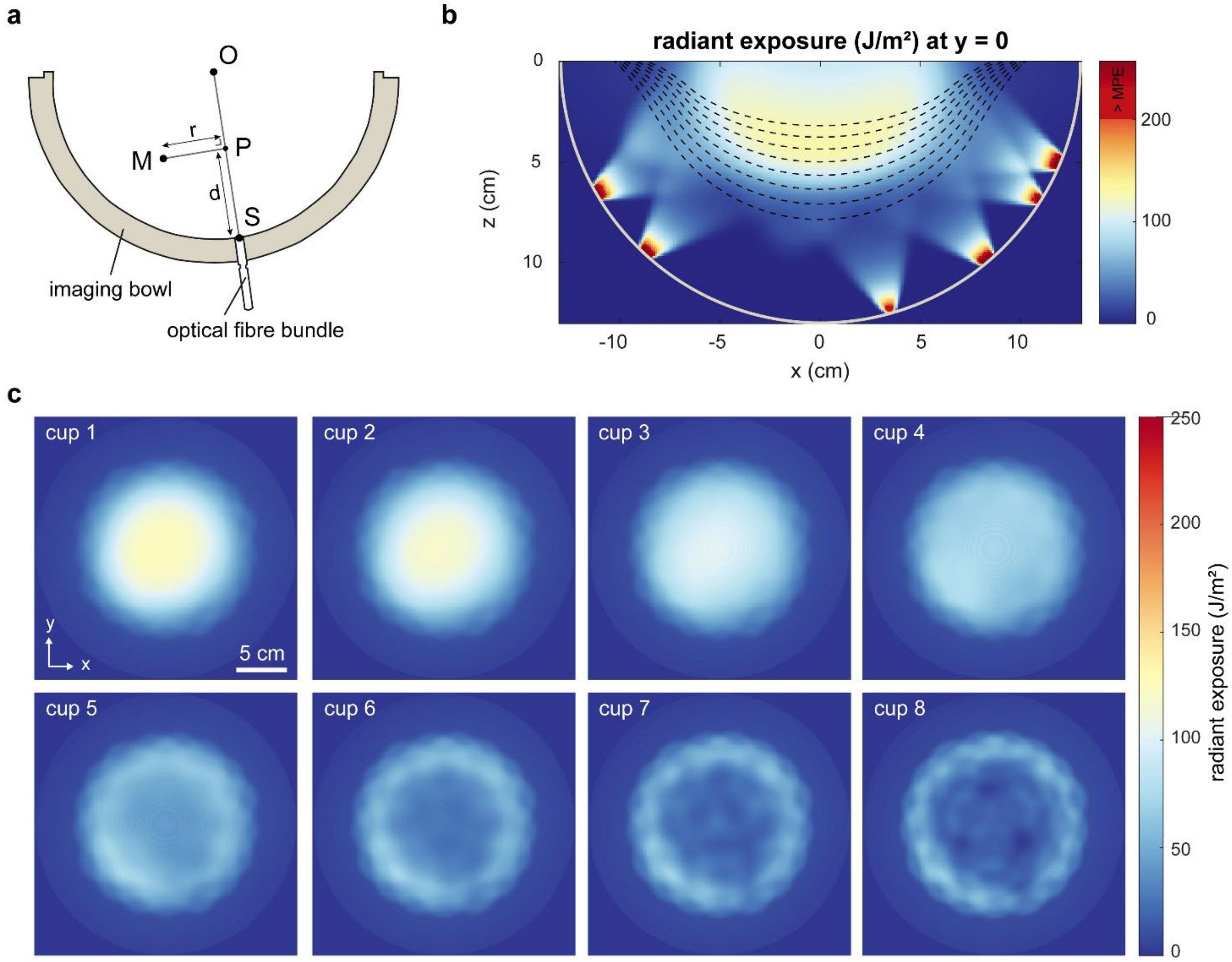

**Figure A2.** (A) Illustration showing the source (S), measurement point (M), origin (O) and the projection point (P) in a cross section of the imaging bowl to clarify the vector model. One optical fibre bundle protruding the bowl is illustrated. (B) The radiant exposure in the imaging bowl at y=0 resulting from our model. The dashed lines show the contours of the eight available cup sizes. All pixels with exposure above the maximal permitted exposure of 200 J/m2 are colored red and show that the exposure within the cups confirms to the safety limits. (C) The radiant exposure on the surface of the eight cups as seen from the bottom of the imaging bowl. Note that the exposures modeled do not incorporate bowl rotations.

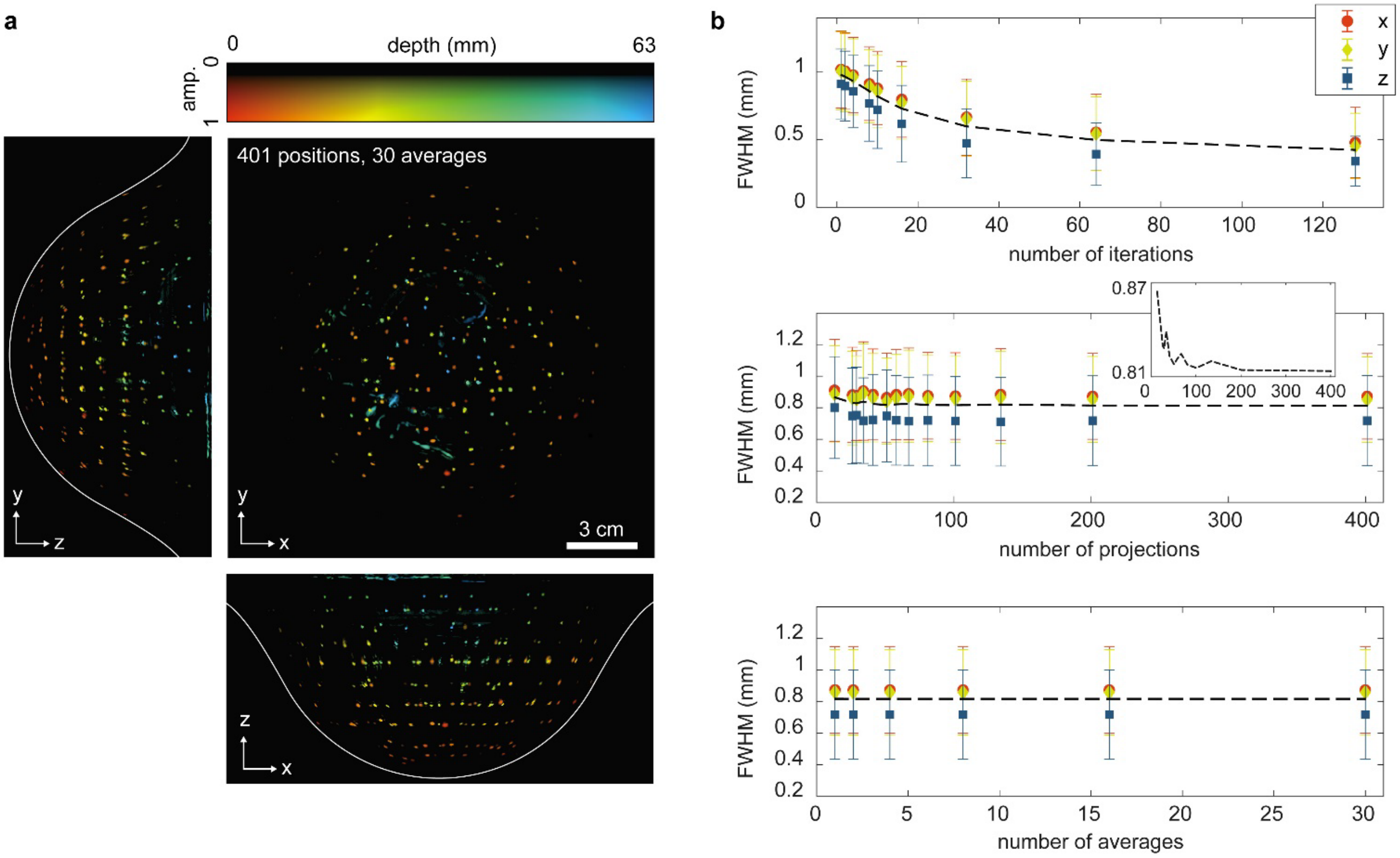

**Figure A3.** Depth color-coded MIPs from an oversampled measurement (800 nm, 401 projections, 30 averages, 10 iterations) on the PSF object. (B) The average FWHM of the Gaussians fitted to 210 reconstructed PSFs plotted for the three reconstruction planes as a function of reconstruction iterations, measurement projections and averages. Error bars indicate standard deviations and the black dashed line is the mean of the FWHM in the three dimensions. The mean curve belonging to the number of projections is shown additionally in an insert to show its decay.

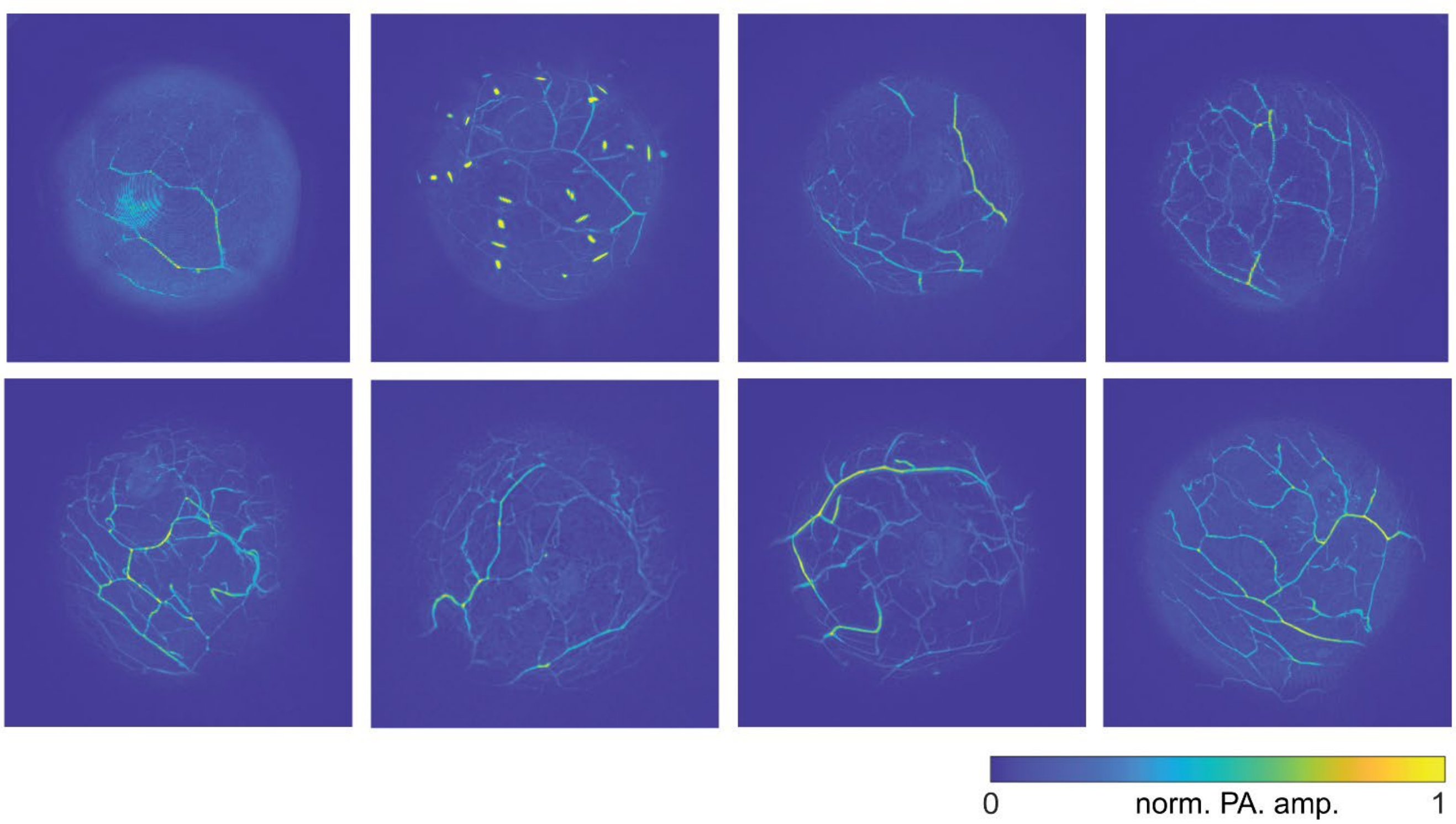

**Figure A4.** PA AP MIPs of one of the breasts of all the volunteers included in the study.

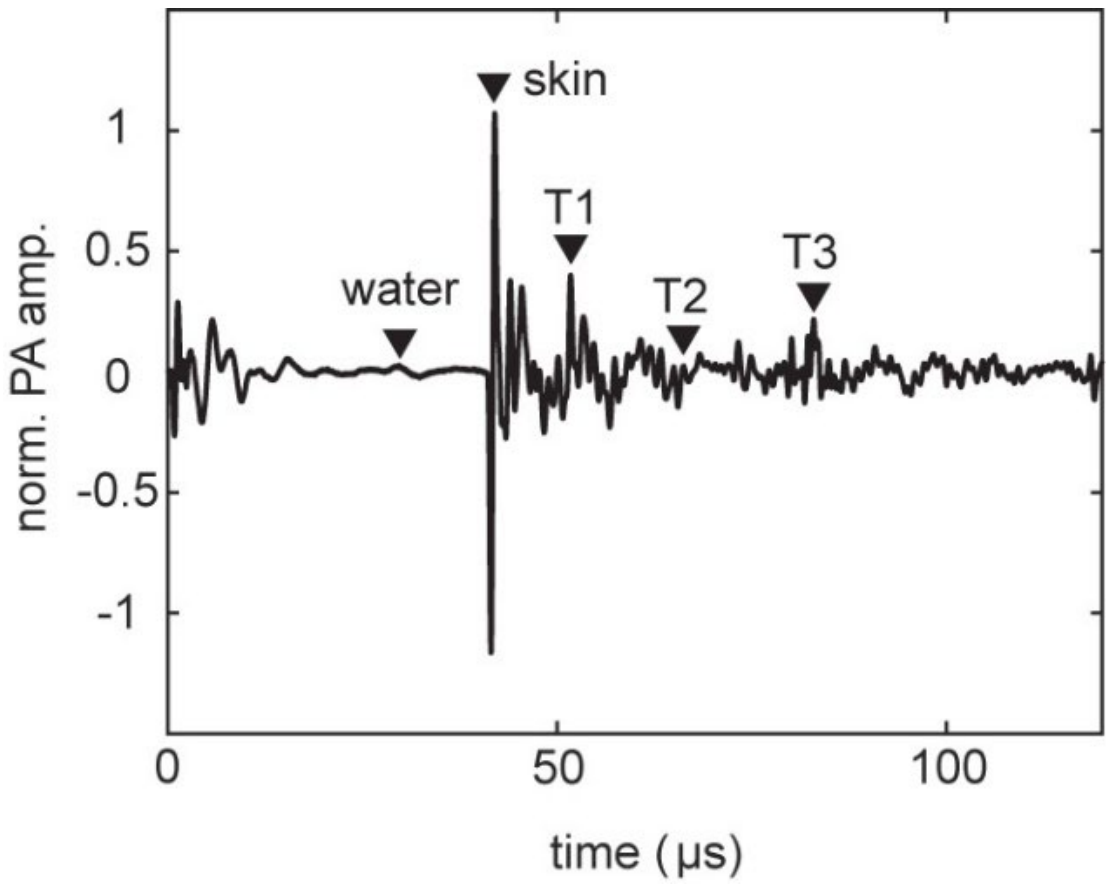

**Figure A5:** PA time trace recorded by the transducer at the bottom at the imaging tank during the measurement on volunteer 016. Peaks that were used to plot the spectra in Figure 5d are highlighted.

**Table A1.** Mean and standard deviations of the radiant exposures on breast surface for different cup-sizes.

| Cup size | Mean (J m$^{-2}$) | Std deviation (J m$^{-2}$) |
|---|---|---|
| 1 | 29.04 | 18.65 |
| 2 | 28.10 | 17.21 |
| 3 | 26.29 | 14.66 |
| 4 | 23.55 | 11.37 |
| 5 | 23.47 | 8.81 |
| 6 | 17.94 | 7.45 |
| 7 | 16.07 | 6.61 |
| 8 | 14.51 | 5.97 |

**Table A2.** Details of image acquisition of all volunteers in the manuscript.

| Volunteer | Wavelengths (nm) | Number of views | Number of averages | Acquisition time (minutes) |
|---|---|---|---|---|
| 1 | 800, 860 | 101 | 12 | 6 |
| 2 | 870 | 101 | 10 | 3 |
| 3 | 720, 755, 797, 833, 860, 890 | 101 | 4 | 6 |
| 4 | 720, 755, 797, 833, 870 | 101 | 4 | 5 |

**Table A3.** Description of Fitzpatrick scale skin types

| | |
|---|---|
| Type 1 | Highly sensitive, always burns, never tans. Example: Red hair with freckles. |
| Type 2 | Very sun sensitive, burns easily, tans minimally. Example: Fair skinned, fair haired Caucasians. |
| Type 3 | Sun sensitive skin, sometimes burns, slowly tans to light brown. Example: Darker Caucasians. |
| Type 4 | Minimally suns sensitive, burns minimally, always tans to moderate brown. Example: Mediterranean type Caucasians, some Hispanics. |
| Type 5 | Sun insensitive skin, rarely burns, tans well. Example: some Hispanics, some Blacks. |
| Type 6 | Sun insensitive, never burns, deeply pigmented. Example: Darker Blacks. |

**Table A4.** Questionnaire used for skin type scaling on the Fitzpatrick scale. Summing the points corresponding to the answers gives the six skin types described in Table A1. I: 0-6 points, II: 6-14 points III: 15-22 points, IV: 22-27 points, V and VI: >27 points

| | 0 | 1 | 2 | 3 | 4 |
|---|---|---|---|---|---|
| What is your eye color? | light blue/ grey/ green | blue/ grey/ green | blue | dark brown | black brown |
| What is your hair color? | red | blond | chestnut/dark blond | dark brown | black |
| How dows your skin react to sunlight? | burns quickly, blisters directly form | burns quickly, blisters after a while | burns sometimes | incidental burns | never burns |
| Does your skin directly tan after exposure to sunlight? | never | rarely | sometimes | often | always |
| How much do you tan? | barely | light colouring | colouring | quickly colouring | quickly turn dark brown |
| What is the tone of your skin that has not been exposed to sunlight? | red | pale | pale with beige tones | brown | dark brown |
| How many birthmarks do you have? | many | some | few | rarely | none |
| How often is your breast directly exposed to sunlight or light from a tanning bed? | never | rarely | sometimes | often | always |
| When was the last time your breasts were exposed to sunlight? | >3 months ago | 2-3 months ago | 1-2 months ago | <1 month ago | <2 weeks ago |

## Additional system information

The hemispherical bowl is made of polymethylmethacrylate (PMMA) with inner surface coated with a 200 nm thick reflecting gold layer. Water inlet and outlets are present to fill and drain the bowl with degassed, demineralized and UV sterilized water for acoustic coupling, maintained at 25 ºC. The imaging bowl together with all its inserts and the I/O modules is connected to a large race wire bearing driven by a stepper motor and a sprocket to allow continuous rotation of the bowl over 360 degrees.

The DAQ consists of 32 pre-amplification channels leading to four 8-channel analog front end (AFE) chips with 14-bit Analog-Digital Converters (ADC). The time traces amplified by 24 dB, are digitized with 50 MHz and 25 MHz sampling frequencies for the PA and US mode respectively, and transmitted to the DAQ-PC. The US Actuation part is responsible for sending the driving signals to the US transducers. It consists of eight 4-channel Digital-Analog Converters (DACs) to send pre-programmed Gaussian-modulated single cycle sinusoidal pulses with a 1 MHz center frequency to the transducers at a 100 Hz repetition rate, where only one transducer emits and all the others receive.

## Laser and light delivery

Both laser heads in the custom-designed high energy tunable laser unit emit pulse-to-pulse wavelength tunable 4.2 ns pulses at 10 Hz repetition rates. In normal operating mode, the pump lasers are triggered with a 90 ns delay between them, to avoid overlap of the two pulses in time in order to stay under the damage threshold of the fiber bundle (see Fig. 1e). In addition to the double pump laser approach, a novel method for compensating depolarization losses is implemented[1] which contributes to achieving high output powers.

The 40 fiber bundles each have a core diameter of 1.67 mm and have a micro lens array beam spreader at the tip providing a beam profile that can be approximated by a sixth-order super-Gaussian. The maximum permitted exposure of 200 $J/m^2$ for the skin at 10 Hz is only exceeded up to a few mm from the fiber bundle surface (see Supplementary Figure A2 for the modeled exposure in the imaging volume). The breast-supporting cups restrict the breast from entering this zone.

## System control and data storage

The measurement sequence can be programmed into the control software by the operator. This allows the number of bowl positions, number of PA excitations per position (averages), US shots per position and the number of PA excitation wavelengths to be varied per measurement. In the example measurement sequence (Figure 1e) for each bowl position, PA measurements are performed at five wavelengths. At each wavelength, two combined light pulses are sent into the bowl; a single combined light pulse comprises the outputs of the two pump lasers (L1 and L2) with an 90 ns delay between them. After each combined light pulse is launched, all 512 detection elements are read out and the measured time-traces stored. Thereafter the bowl rotates to a new position, to repeat the measurement sequence. The repetition rate of the laser is 10 Hz, while that of the US emission is 100 Hz. This permits 9 US measurements to be performed in the interval between two PA measurements.

After the operator has selected the sequence, the operator-PC generates a measurement program which is then automatically programmed on the I/O modules and the trigger module. Once the measurement has started, the recorded PA and US time traces are amplified, digitized and stored on the DAQ-PC. The total recorded data size of a measurement depends on the chosen measurement sequence but is usually in the order of 65 GB for a multi-wavelength tomographic PA-US measurement. After the measurement has finished, all the data is downloaded from the DAQ-PC and uploaded to the image reconstruction PCs for generation of the images.

## Spatial resolution calculation

The spatial resolution of the PAM3 system was assessed from the reconstructed images of the over-sampled PA measurement on the PSF test object, consisting out of 401 bowl rotation steps and 30 averages. Reconstructions were made with all the recorded time traces but also with down sampled sets of the recorded data to investigate the effect of the number of bowl positions and averages on the spatial resolution (see Supplementary Figure A2). No post-processing was applied to the reconstructed images. The spatial coordinates of the centers of all the 213 beads in the test object were manually determined based on an initial reconstruction. Then, for each bead, the nearest local intensity maximum was found around which a ROI of 2x2x2 $mm^3$ was selected. In the ROI, Gaussians were fitted to the intensity profiles crossing the beads in *x*, *y* and *z* direction. The full width at half maximum (FWHM) of these Gaussians was then calculated as a value for the spatial resolution. Beads that were closer than 3 mm to another bead were discarded, as these may induce inaccuracies in the Gaussian fitting.

### Computation time

In the current implementation, a single wave simulation takes 10 minutes on a TITAN RTX GPU (see Section "Photoacoustic inversion"), and each iteration requires two of such simulations. As a result, 10 iterations take 3h 20min and reconstructing a data set with 5 wavelengths can take up to 20h (including pre-computations).

### References

1. A. Michalovas, Depolarization compensator, EU patent EP3712664 (A1), 2020-09-23

********